\documentclass[11pt,twocolumn]{article}

\usepackage[utf8]{inputenc}
\usepackage[T1]{fontenc}
\usepackage{microtype}
\usepackage{amsmath,amssymb}
\usepackage{graphicx}
\usepackage{booktabs}
\usepackage{setspace}
\usepackage{url}
\usepackage[hidelinks]{hyperref}
\usepackage{authblk}
\usepackage[margin=0.75in]{geometry}
\usepackage{authblk}
\usepackage[font=small,labelfont=bf]{caption}
\usepackage[numbers,sort&compress]{natbib}
\usepackage{placeins}

\title{\textbf{Human--LLM Collaboration Is Transforming Complexity Metrics in Scientific Texts}}

\author[1,2]{R. Alexander Bentley}
\author[3,1]{Blai Vidiella}
\author[4,1]{Damian J. Ruck}
\author[5,1]{Senjuti Dutta}
\author[6,1]{Kai Li}
\author[7,8,1]{Sergi Valverde}

\affil[1]{Center for the Dynamics of Social Complexity (DySoC), University of Tennessee, Knoxville, TN 37996, USA}
\affil[2]{Department of Anthropology, University of Tennessee, Knoxville, TN 37996, USA}
\affil[3]{Centre for Biodiversity Theory and Modelling, Theoretical and Experimental Ecology Station, CNRS, Moulis, France}
\affil[4]{Advai Ltd., 20--22 Wenlock Road, London N1 7GU, UK}
\affil[5]{Institute of Cognitive Science, University of Colorado Boulder, Boulder, CO 80309, USA}
\affil[6]{School of Information Sciences, University of Tennessee, Knoxville, TN 37996, USA}
\affil[7]{Institute of Evolutionary Biology (CSIC--UPF), Pg. Barceloneta 37, Barcelona 08003, Spain}
\affil[8]{European Centre for Living Technology (ECLT), Ca' Bottacin, Dorsoduro 3911, 30123 Venice, Italy}

\date{}

\begin{document}

\maketitle

\begin{abstract}
\footnotesize 
\footnotesize
\begin{spacing}{1}
While human language has long been studied as a complex system, Large Language Models (LLMs) are rapidly becoming contributors to its dynamics. Because LLMs are trained on human language use, their effects on the broader human--AI linguistic ecosystem are likely subtle at first. As their use becomes more widespread, however, LLMs may alter emergent properties of language, particularly as models increasingly train on mixed human--LLM textual data. Here, we draw on complexity science to look for subtle LLM effects in millions of arXiv abstracts from 2010 to 2025. The year 2023, when LLMs rapidly became widely used, serves as a landmark in a natural experiment. While we find a sharp increase in a composite LLM-associated style index after early 2023, we observe only subtle changes in the exponents of Zipf's law and Heaps' law. More compelling, however, are two subtle changes in complexity metrics that emerge from 2023 onward. First, turnover among top-ranked words increases sharply. Second, the positive relationship between the LLM-associated style index and three complexity metrics---vocabulary size and the exponents of Heaps' and Zipf's laws---becomes flatter after 2022. Together, these patterns are consistent with changes in the emergent properties of scientific text in a mixed human--AI linguistic ecosystem.
\end{spacing}
\end{abstract}


\section*{Introduction}

Large Language Models (LLMs), rapidly adopted since early 2023, are transforming education, scientific research, and cultural communication, reshaping how knowledge is produced, evaluated, and transmitted. As LLMs support tasks ranging from literature reviews and data analysis to collaborative reasoning \cite{Wang_etal_2023, Yerramilli-Rao_etal_2025, Zenil_etal_2023, King_etal_2009, Skarlinski_etal_2024}, surveys indicate that a substantial fraction of users also engage with LLMs for psychological support \cite{Rousmaniere_etal_2025}. As human trust in scientific expertise evolves \cite{Druckman_etal_2025}, dyadic conversations with an LLM can reduce conspiracy beliefs \cite{Costello_etal_2024}, and in group settings they guide diverse participants toward consensus \cite{Tessler_etal_2024}.

As LLMs become embedded in daily communication, humans and models increasingly form a mixed linguistic population \cite{Ashery_etal_2025}. There is bidirectional learning: LLMs are trained on human texts \cite{Gupta&Pruthi_2025}, while human language adapts to LLM outputs \cite{Keener_2025, Yakura_etal_2024}. This reciprocal coupling makes it possible to treat LLMs as a new feedback within cultural evolution \cite{Ashery_etal_2025, Perez_etal_2024b, Brinkmann_etal_2023, Perez_etal_2024, Suzuki&Arita_2023}, potentially as consequential as early writing, markets, or early governments \cite{Farrell_etal_2025}. Unlike Deep-Q systems of the 2010s, which learned through largely solitary reinforcement learning \cite{LeCun_etal_2015}, LLMs generate responses by sampling from probability distributions learned from large corpora of human-generated text \cite{Evans_etal_2026}. 

What will be the future of knowledge or language in this increasingly mixed human-AI population? There are at least two perspectives. On one hand, LLMs may reduce lexical complexity, partly through sentence simplification \cite{Qiang_etal_2025} but also through recycling their training data with limited novelty/invention \cite{Bentley_etal_2004,Duran-Nebreda_etal_2022, RosilloRodes_etal_2026, Vidiella_etal_2022}. Additionally, if AI-generated content begins to overwhelm human-authored content in AI model training data \cite{Jones_2024, Longpré_etal_2024}, successive generations of AI models may increasingly be trained on their own outputs, potentially diminishing quality and complexity \cite{Peterson_2024, Shumailov_etal_2024}. On the other hand, interacting LLM systems may produce emergent dynamics that cannot be reduced to individual models \cite{Zomer&DeDomenico_2026}. Collective biases can arise that are not simply the average of individual tendencies, with outcomes shaped by feedback, coordination, and the formation of shared conventions \cite{Ashery_etal_2025}. In this view, collective LLM outputs are better understood as emergent outcomes of complex interactions between pre-trained knowledge, embedded priors, and social context \cite{Ferrarotti_etal_2026}.

To connect LLM text generation with models of cultural evolution, we begin with the core sampling process. In  LLMs, the probability, $P_i$, of choosing token $i$ essentially follows a softmax function involving logit scores $z_i$ of the tokens and temperature $T$:
\begin{equation}
P_i = \frac{e^{z_i / T}}{\sum_{j=1}^{N} e^{z_j / T}}
\label{eq:LLM}
\end{equation}

This is similar to a general model of cultural evolution \cite{Bentley_etal_2014, Brock_etal_2014, Vidiella_etal_2022}, in which the probability of choosing token $i$ at time $t$ is:
\begin{equation}
P_i = p_i(t)^J \frac{ e^{\beta U_i}}{\sum_{j=1}^{N(t)} p_j(t)^J e^{\beta U_j}}
\label{eq:Vidiella}
\end{equation}

\noindent where $p_i(t)$ is the relative frequency of token $i$ among the $N(t)$ different tokens at time $t$, $U_i$ is its utility, $J$ is social learning intensity and $\beta$ is transparency of utility \cite{Bentley_etal_2014, Vidiella_etal_2022}. Equation \ref{eq:Vidiella} captures realms of complexity, including rational individual choice (high $\beta$, low $J$), following experts (high $\beta$, high $J$) and random choice ($J = 0$ and $\beta = 0$), with a rich diversity of dynamics from this simple model \cite{Vidiella_etal_2022}. 

When $J = 1$ and $\beta = 0$ in Equation \ref{eq:Vidiella}, the process should resemble a neutral copying model \cite{Bentley_etal_2004}, where choices are made with probabilities proportional to frequency $p_i$. An important difference with LLMs in equation (\ref{eq:Vidiella}) is the $p_i^J$ term, which captures the popularity bias.  In contrast, $z_i/T$ in Equation (\ref{eq:LLM}) does not directly weight choices by token frequency. At $T=1$, the LLM samples from its native learned distribution, while increasing $T$ progressively flattens differences among token logits. Using a first-order expansion for large $T$,
\begin{equation}
P_i \approx \frac{1}{N} + \frac{1}{NT}(z_i-\bar z),
\label{eq:LargeT}
\end{equation}
where $\bar z$ is the mean logit across candidate tokens, showing that as $T \to \infty$ the distribution approaches uniform random choice, $P_i \to 1/N$, analogous to lowering $\beta$ in Equation \ref{eq:Vidiella} with $J=0$ (not $J=1$). Because randomness in LLM outputs arises from sampling procedures rather than individual stochastic choice, it may be that noisy LLM input into the lexical ecosystem increases turnover in word rankings. 

Introducing LLMs among humans as a mixed population, then, could either increase or decrease randomness, depending on whether LLM outputs amplify conformity or inject novelty. A comprehensive survey of the linguistic characteristics of AI-generated text found that a large number of studies converge on the conclusion that AI-generated text is generally less lexically diverse than human-written text, also exhibiting a smaller vocabulary size, higher repetition rates, and higher n-gram reuse.  A comparison of human versus LLM-generated news text found that human texts consistently exhibited higher lexical diversity than all LLM families tested \cite{MuozOrtiz_etal_2023}.  Another study, however, found greater lexical diversity in LLM-generated text than human student-written texts \cite{Reinhart_etal_2025}.

This suggests a range of predictions for the stochastic process of word selection, drawing on iterated simulations of the unified cultural evolution model \cite{Vidiella_etal_2022} as well as neutral copying models \cite{Bentley_etal_2004, Gleeson_etal_2013, Duran-Nebreda_etal_2022}. These models generate several empirical signatures, including the well-known Zipf distribution of variant frequencies and Heaps’ law between corpus size and the number of unique tokens, and turnover in the top-$y$ ranked items, denoted $z_y$, measured as the fraction of new items entering a ranked list of the top y items at each time step \cite{Bentley_etal_2007, Acerbi&Bentley_2014, Ruck_etal_2017}. 

In neutral models, turnover $z_y \propto \mu^{0.55} y^{0.86} N^{0.1}$, where $N$ is population size, invention rate $\mu$ is the proportion of unique tokens introduced per iteration and $y$ is small relative to the total number of variants \cite{Evans&Giometto_2011, Acerbi&Bentley_2014}. Plotting turnover $z_y$ against list size $y$ therefore provides a diagnostic for transmission bias: scaling exponents $b>0.86$ imply conformity bias, $b=0.86$ is consistent with neutrality, and $b<0.86$ suggests pro-novelty bias \cite{Acerbi&Bentley_2014}. Although individual-level mechanisms cannot be inferred from aggregated data, shifts in the transparency and social-learning parameters of mixed human–AI populations could change this measurable turnover metric.

To investigate how LLMs affect lexical complexity, we exploit a natural observational experiment beginning in early 2023 with widespread use of LLMs. ChatGPT was released publicly in November 2022 and experienced rapid adoption, reaching tens of millions of users within weeks and approximately 100 million by early 2023 \cite{Burmagina_2026}. By early 2023, it is estimated that over a third of computer science arXiv abstracts had been ChatGPT-revised \cite{Geng&Trotta_2024, Kobak_etal_2025}. With continued growth thereafter, we treat 2023 as the first year of widespread LLM input into online texts. We analyze millions of arXiv paper abstracts from 2010 to 2025 to assess changes from 2023 onward in three measures of lexical structure: vocabulary size, Heaps’ law exponent, Zipf’s law exponent and turnover in top-$y$ lists. While turnover charcterizes popularity bias, Heaps’ and Zipf’s laws capture vocabulary growth and lexical concentration, respectively.

\section*{Results}

\begin{figure*}[t]
\centering
    \includegraphics[width=0.3\linewidth]{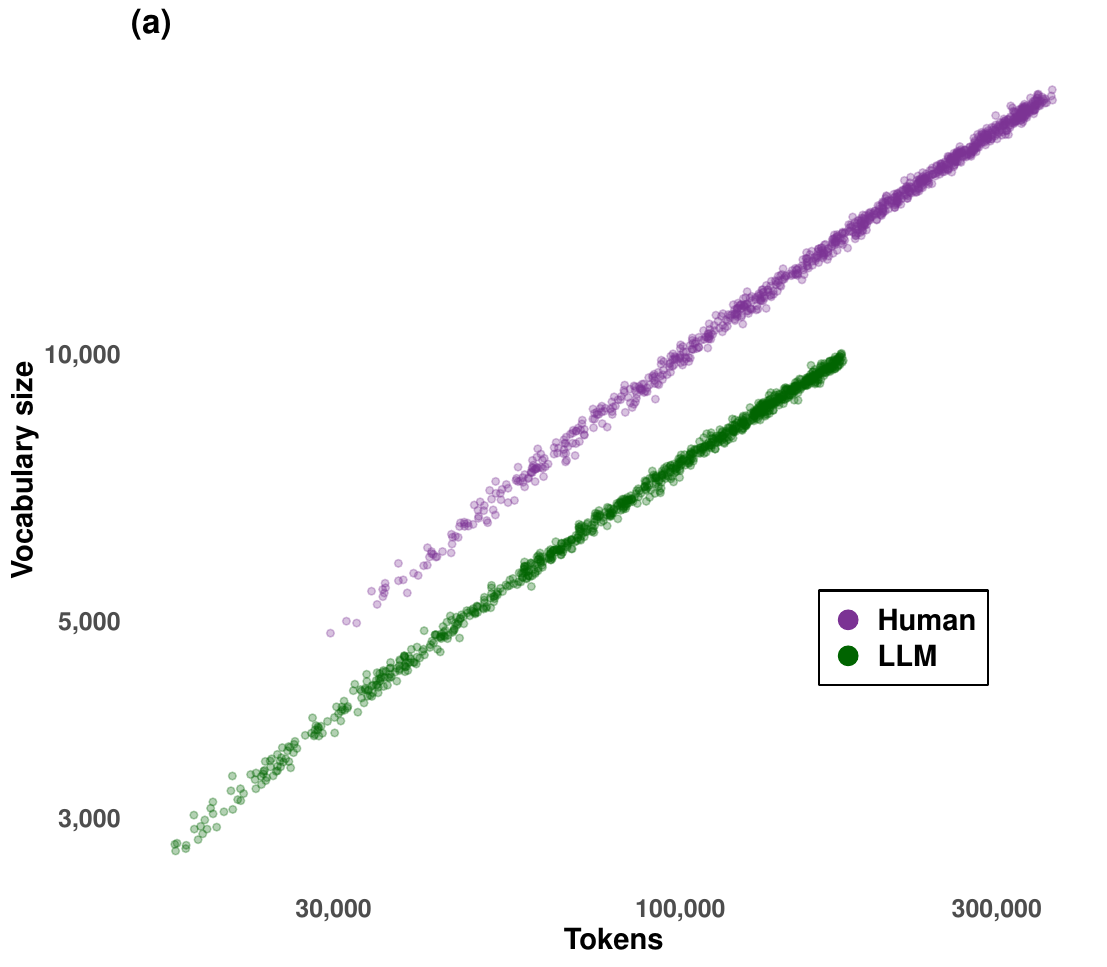}
    \includegraphics[width=0.3\linewidth]{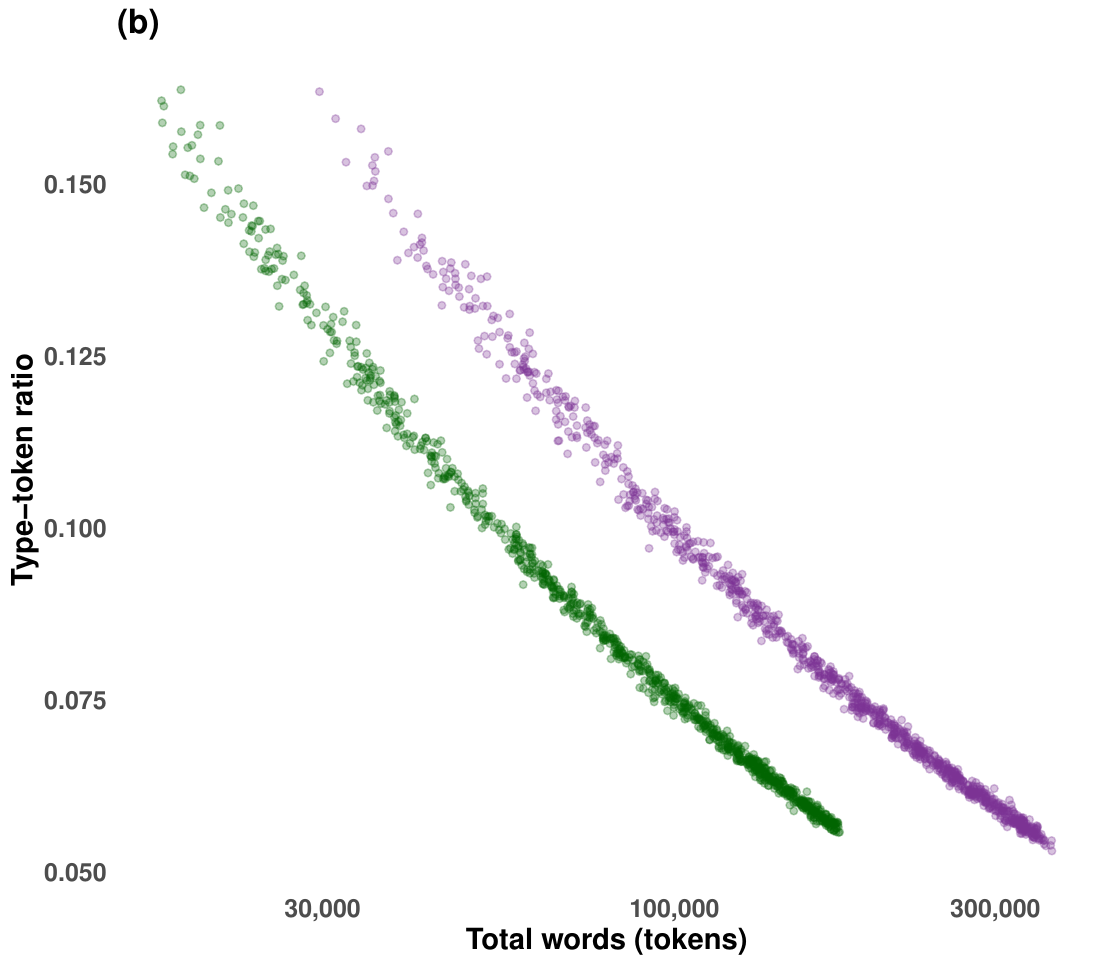}
    \includegraphics[width=0.3\linewidth]{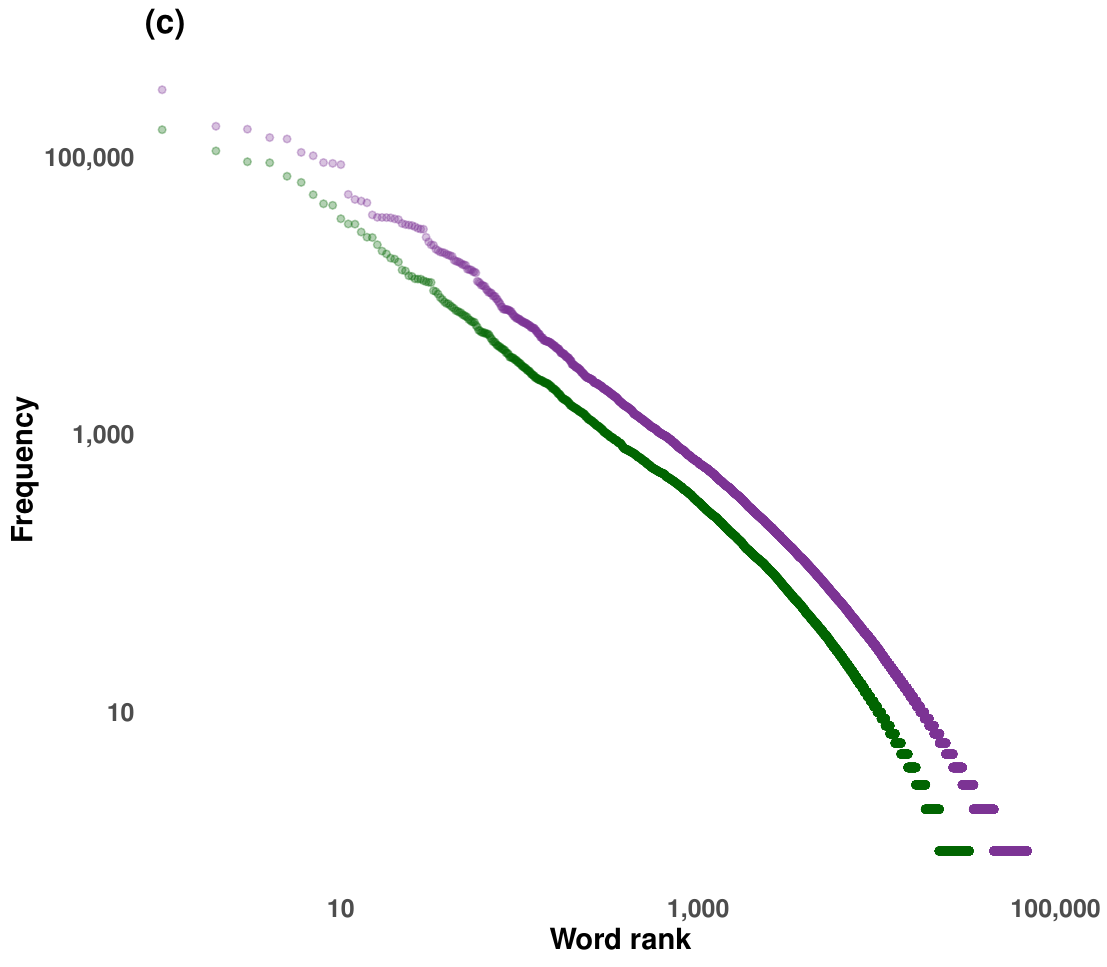}
\caption{Patterns in the HC3 corpus \cite{Gao_etal_2023} comparing human text (purple) and LLM-generated text (green): \textbf{(a)} Heaps’ law (total tokens vs.\ vocabulary size), \textbf{(b)} lexical diversity measured by the type–token ratio, and \textbf{(c)} Zipf’s law.}
    \label{fig:HP3_corpus}
\end{figure*}

To contextualize the observational results, we first compare emergent lexical structure in human- and LLM-generated text (GPT-3.5) using the HC3 corpus \cite{Gao_etal_2023} (Figure \ref{fig:HP3_corpus}). Across Heaps’ law (a), type–token ratio (b), and Zipf’s law (c), LLM-generated texts exhibit a slightly lower Heaps exponent ($b=0.522$ [0.520, 0.524]) than humans ($b=0.539$ [0.537, 0.541]), and a slightly higher Zipf exponent ($\alpha=1.741$ [1.737, 1.745]) than humans ($\alpha=1.681$ [1.679, 1.683]). This similarity is expected, given that LLMs are trained on human texts, and "it has become difficult to distinguish their output from human-written text" \cite{Reinhart_etal_2025}.

\begin{figure*}[h]
    \centering
          \includegraphics[width=0.45\linewidth]{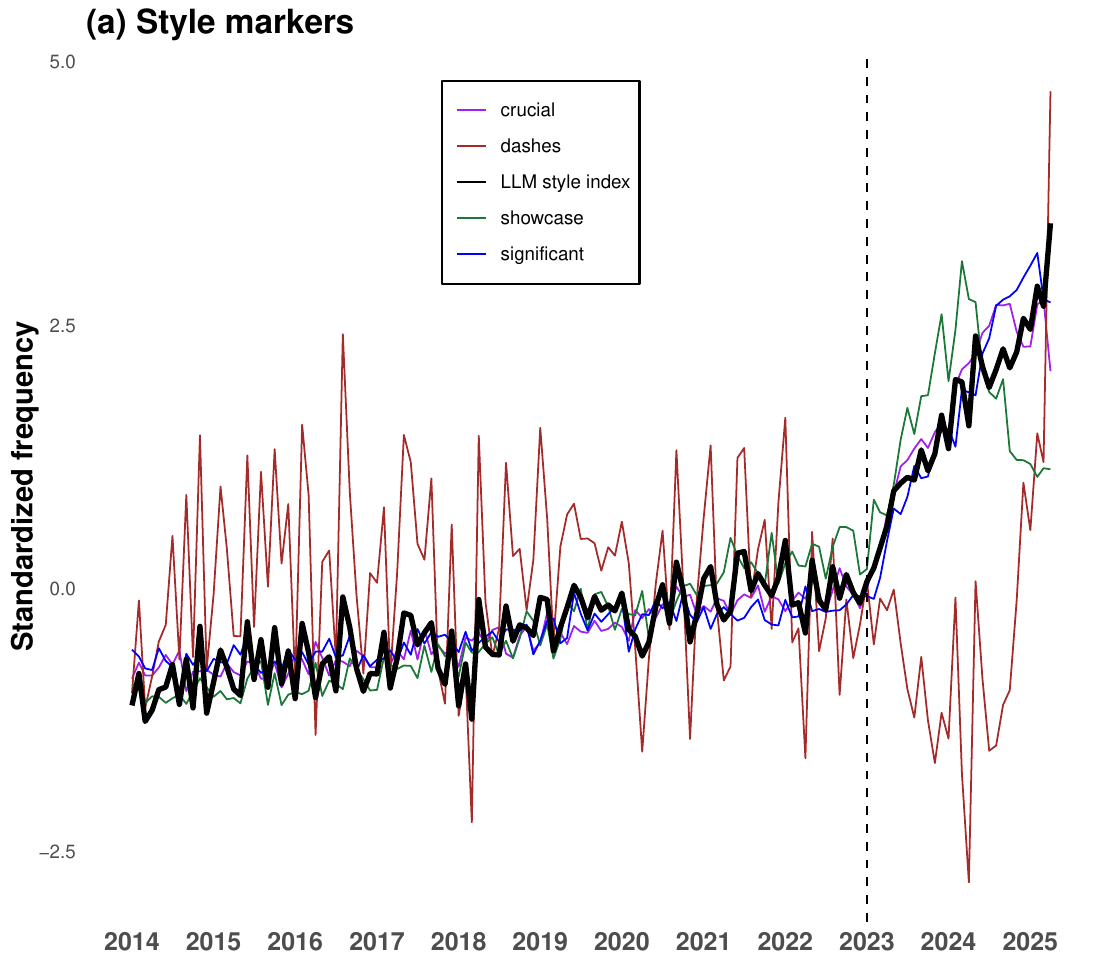}
               \includegraphics[width=0.45\linewidth]{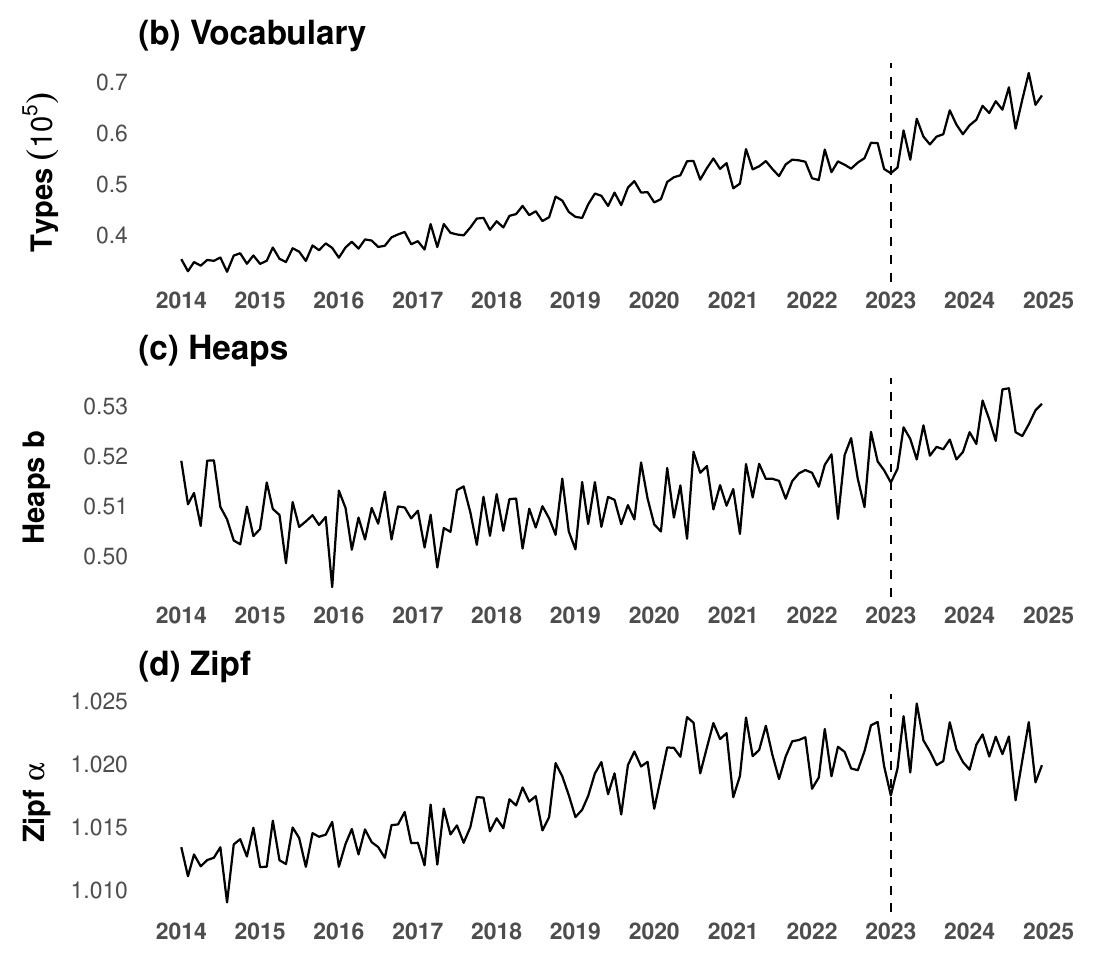}
    \caption{Monthly trends in arXiv abstracts, 2014–2025, showing (a) standardized frequencies (z-scores) of putative markers of LLM influence, including dash usage, selected lexical markers, and a composite LLM-associated style index; (b) vocabulary size; (c) Heaps’ law exponent $b$; and (d) Zipf’s law exponent $\alpha$. Vertical dashed lines mark 2023.
}
    \label{fig:Exponents}
\end{figure*}

Next, we examine temporal patterns in the arXiv abstracts. Figure \ref{fig:Exponents}a shows monthly frequencies, standardized as $z=(x-\mu)/\sigma$, of putative markers associated with LLM use, including dash usage, selected lexical markers, and the composite LLM-associated style index. All of their frequencies rise sharply beginning in 2023, coinciding with the widespread adoption of generative AI tools. 

The right side of Figure \ref{fig:Exponents} shows monthly statistics derived from arXiv abstracts, including (b) vocabulary size, (c) Heaps’ law exponent $b$, and (d) Zipf’s law exponent $\alpha$. Changes around early 2023 are subtle: vocabulary size and the Heaps exponent show modestly accelerated growth, while the Zipf exponent changes comparatively little.

\begin{figure*}[ht]
    \centering
        \includegraphics[width=0.3\linewidth]{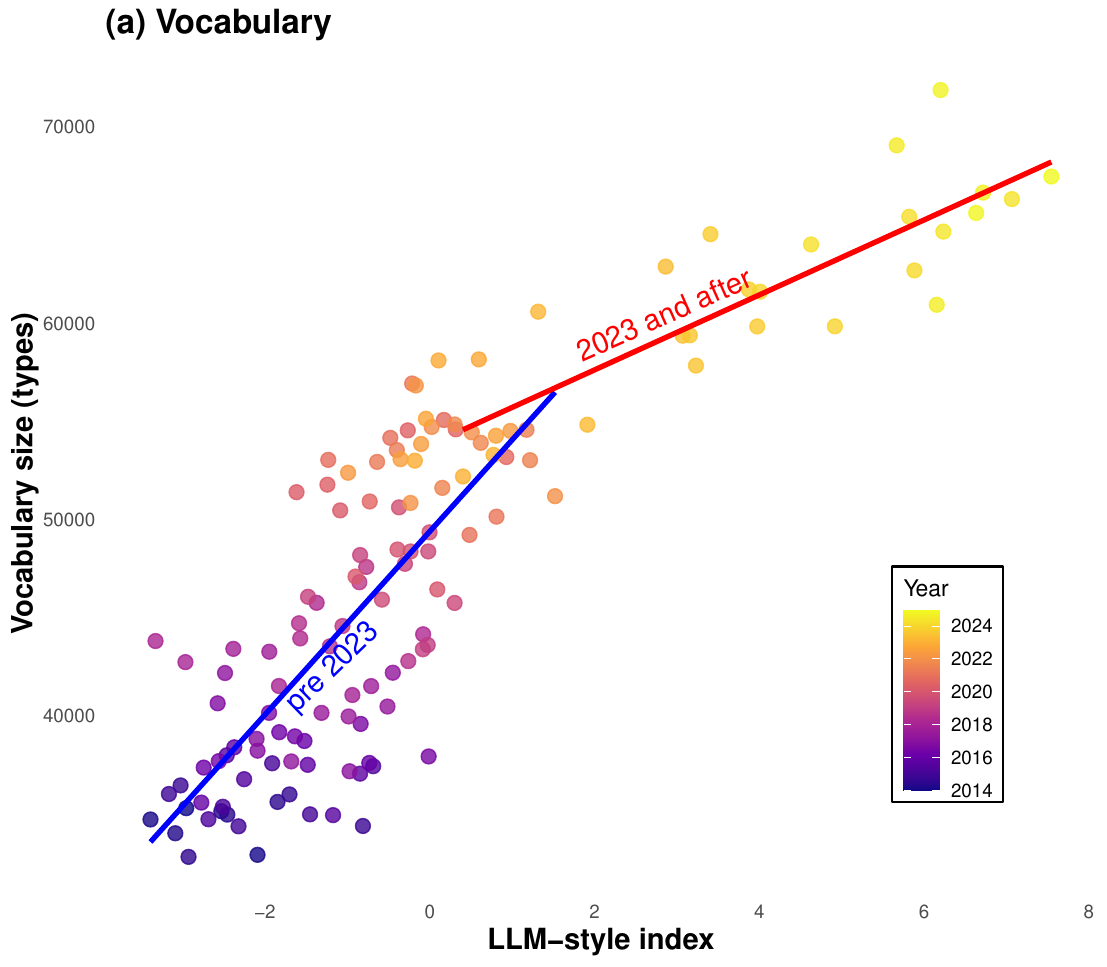}
\includegraphics[width=0.3\linewidth]{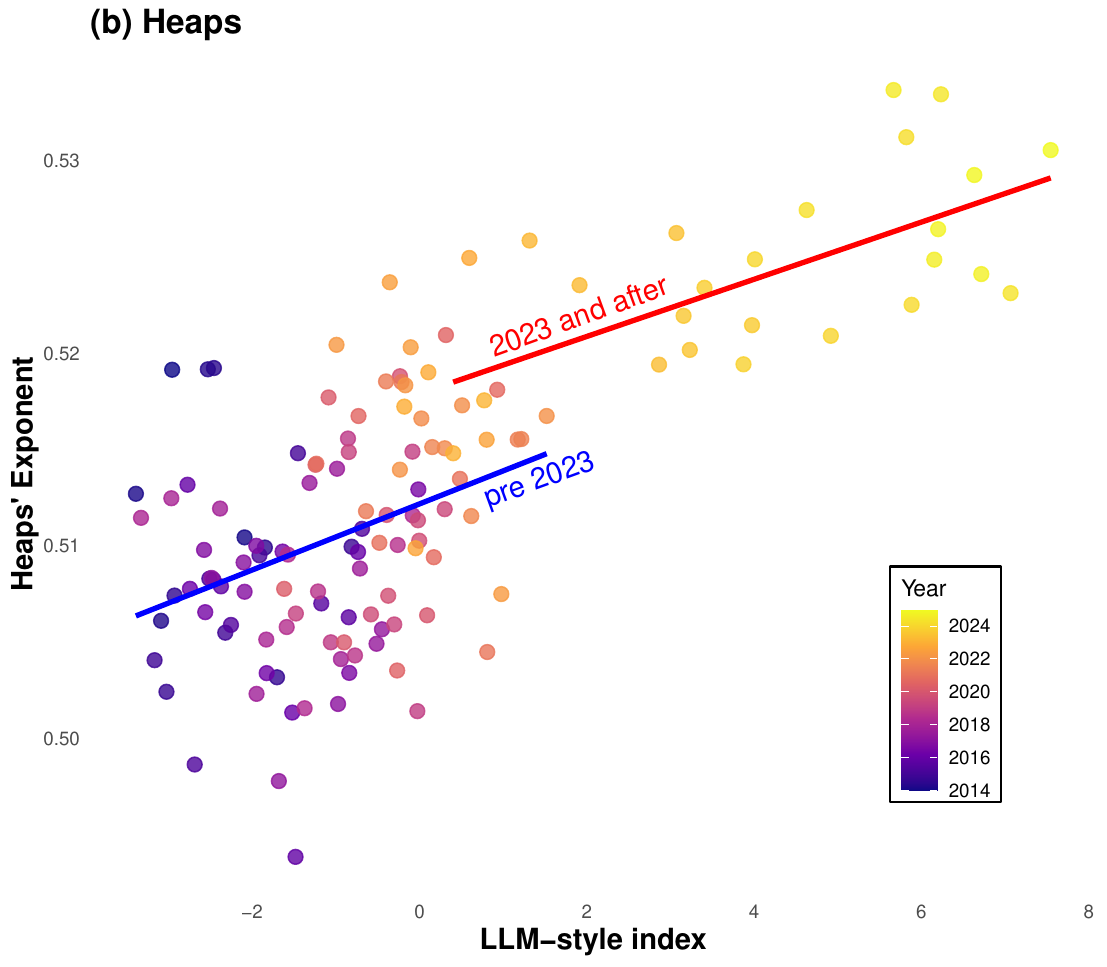}
    \includegraphics[width=0.3\linewidth]{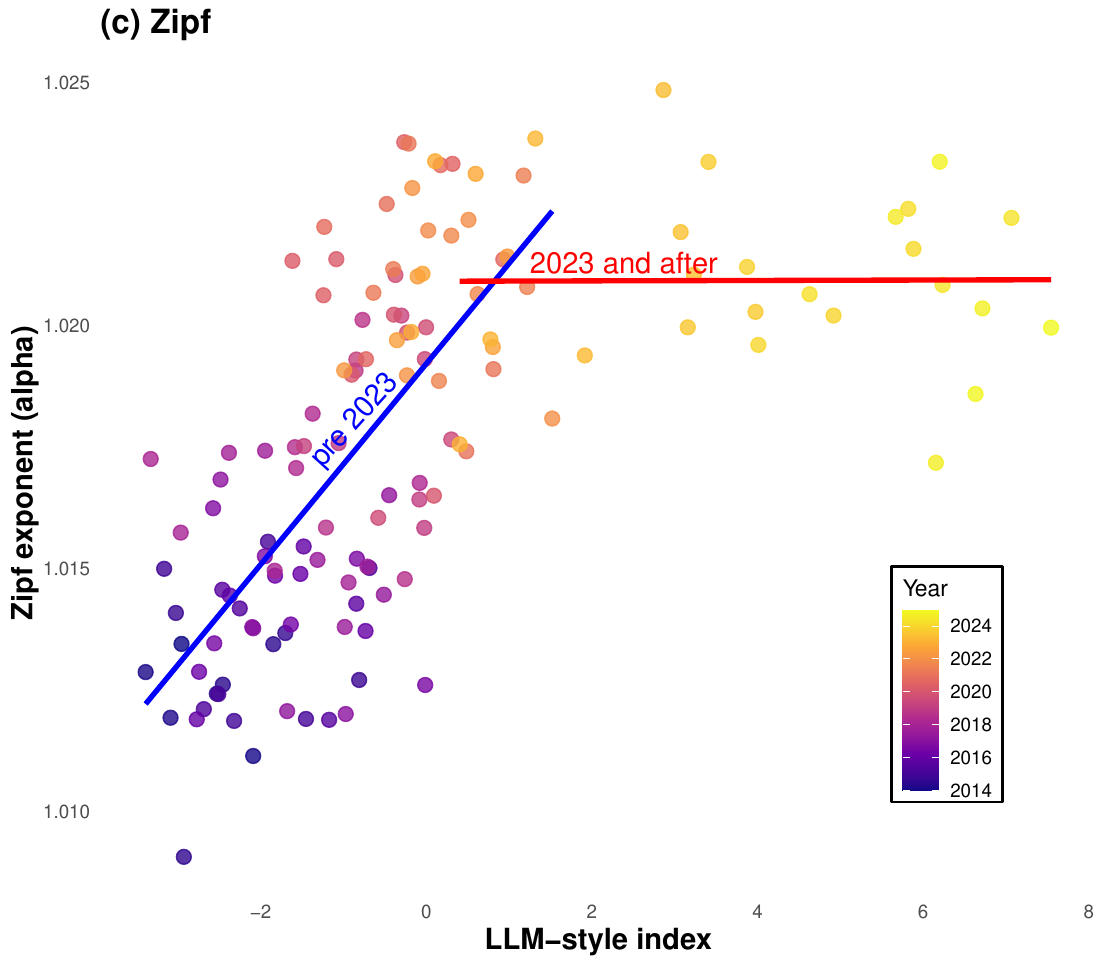}
 \caption{Relationships between a composite LLM-associated style index and measures of lexical structure: (a) vocabulary size (types), (b) Heaps’ law exponent $b$, and (c) Zipf exponent $\alpha$. Each point represents a month (2014–2024), colored by time, with linear fits shown. Blue lines fit data before 2023; red lines 2023 and after.}
    \label{fig:LLM_index}
\end{figure*}

Zooming in on associations with the LLM-associated style index reveals clearer shifts beginning in 2023. Figure \ref{fig:LLM_index} compares the index before and after this point against three measures of lexical structure. Higher index values are associated with larger vocabularies, higher Heaps’ exponents, and increasing Zipf exponents, indicating a pattern in which lexical diversity expands alongside greater concentration in usage patterns. Each point represents a month (2014–2024), colored by time, with separate linear fits for pre-2023 data (blue) and 2023 onward (red). In vocabulary size and Zipf’s exponent, these relationships shift noticeably after 2023, coincident with widespread LLM adoption. 

\begin{figure}[ht]
\centering
\includegraphics[width=0.85\linewidth]{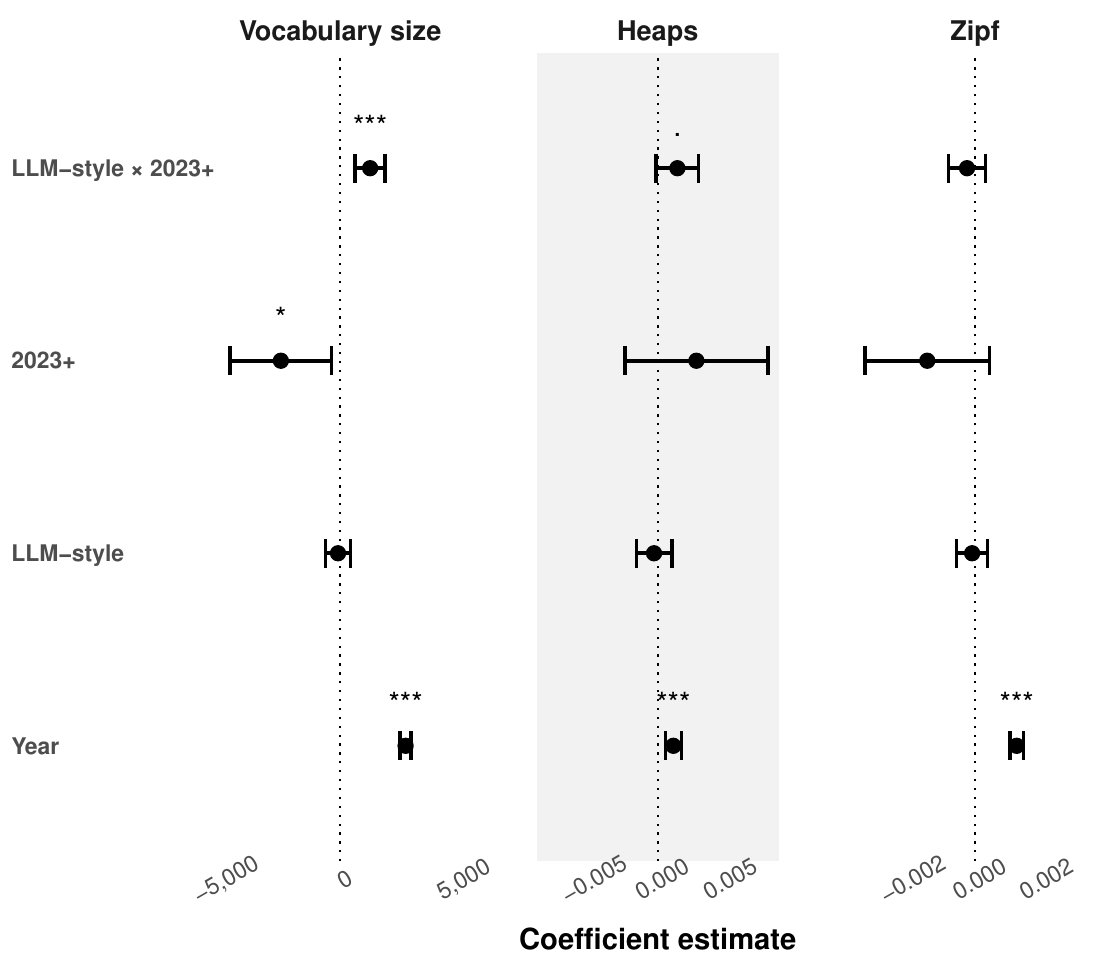}
\includegraphics[width=0.85\linewidth]{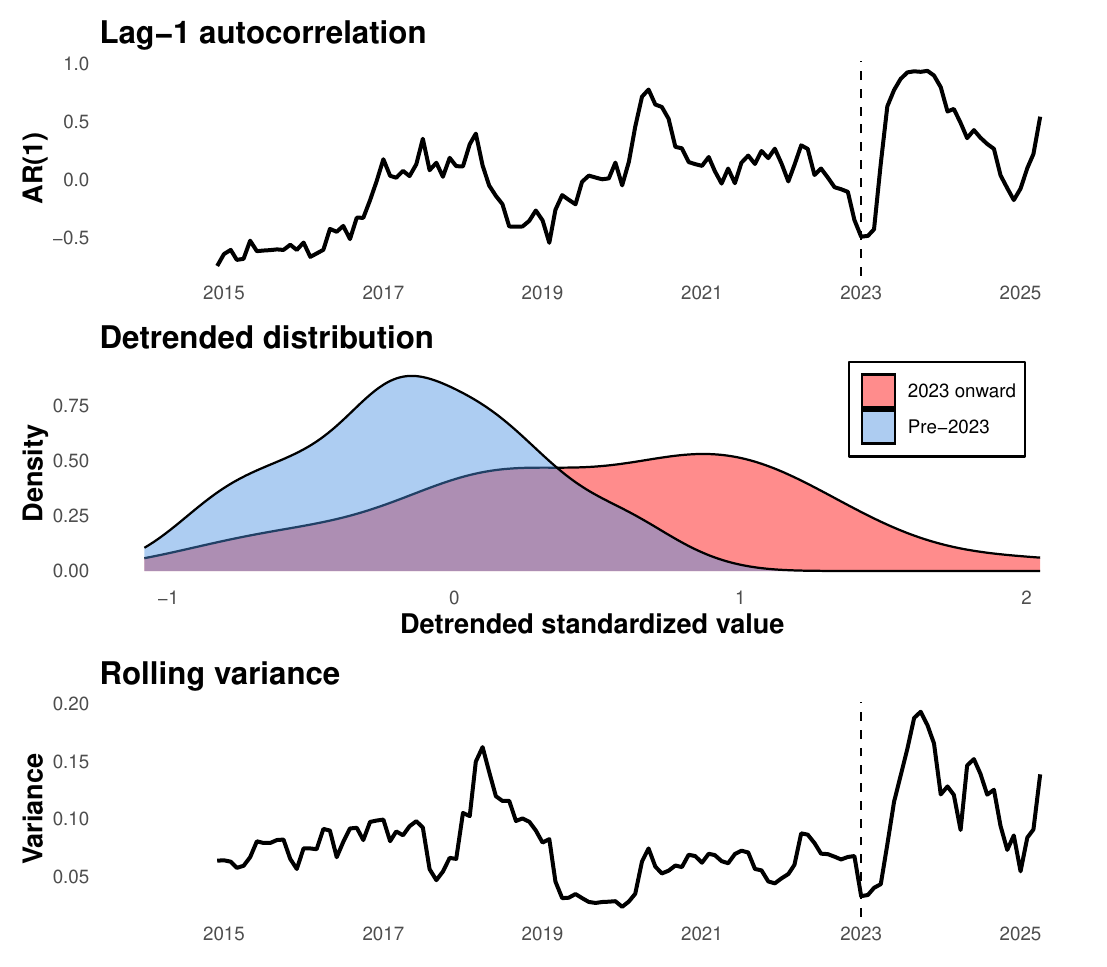}
\caption{Top: Regression coefficients from multivariate models predicting vocabulary size (left), Heaps’ law exponent $b$ (center), and Zipf exponent $\alpha$ (right). Vertical dotted line marks zero effect. Significance levels: *** $p<0.001$, ** $p<0.01$, * $p<0.05$, . $p<0.10$. Bottom: Early warning signal (EWS) diagnostics for the composite LLM-associated style index, including (a) Rolling lag-1 autocorrelation; (b) Distribution of detrended values; and  (c) Rolling variance (12-month window).}
\label{fig:Regressions_EWS}
\end{figure}

Because the LLM-associated style index was increasing through time, as were all three lexical statistics (Figure \ref{fig:Exponents}), it is important to control for publication year when evaluating these positive correlations. The left side of Figure \ref{fig:Regressions_EWS} shows regression coefficients from multivariate models predicting vocabulary size, Heaps’ law exponent $b$, and Zipf exponent $\alpha$ from publication year, a composite LLM-associated style index, a post-2023 indicator, and their interaction. The interaction term tests whether the association between the LLM-style index and each lexical complexity measure changed after 2023 while accounting for long-term temporal trends. A significant positive interaction for vocabulary size, and a weaker positive interaction for Heaps’ exponent (Figure \ref{fig:Regressions_EWS}, left), suggest that the post-2023 shifts are not explained solely by secular temporal trends.

These interaction effects are more diagnostic of the post-2023 changes than standard early warning signal (EWS) diagnostics applied to this time series \cite{Scheffer_etal_2009, Scheffer_etal_2012}. Although the LLM-associated style index increases sharply from 2023 (Figure \ref{fig:Exponents}a), the EWS indicators are comparatively modest (Figure \ref{fig:Regressions_EWS}, right). The rolling lag-1 autocorrelation (12-month window) of the detrended series shows some increase, consistent with critical slowing down. The distribution of detrended values before and after 2023 shows only minor asymmetry, with skewness shifting from 0.028 (pre-2023) to $-0.042$ (2023 onward). As a proxy for “flickering” \cite{Scheffer_etal_2009}, the rolling variance (12-month window) increases after early 2023, but by an amount comparable to a previous increase around 2018.

\begin{figure}[ht]
    \centering
        \includegraphics[width=0.85\linewidth]{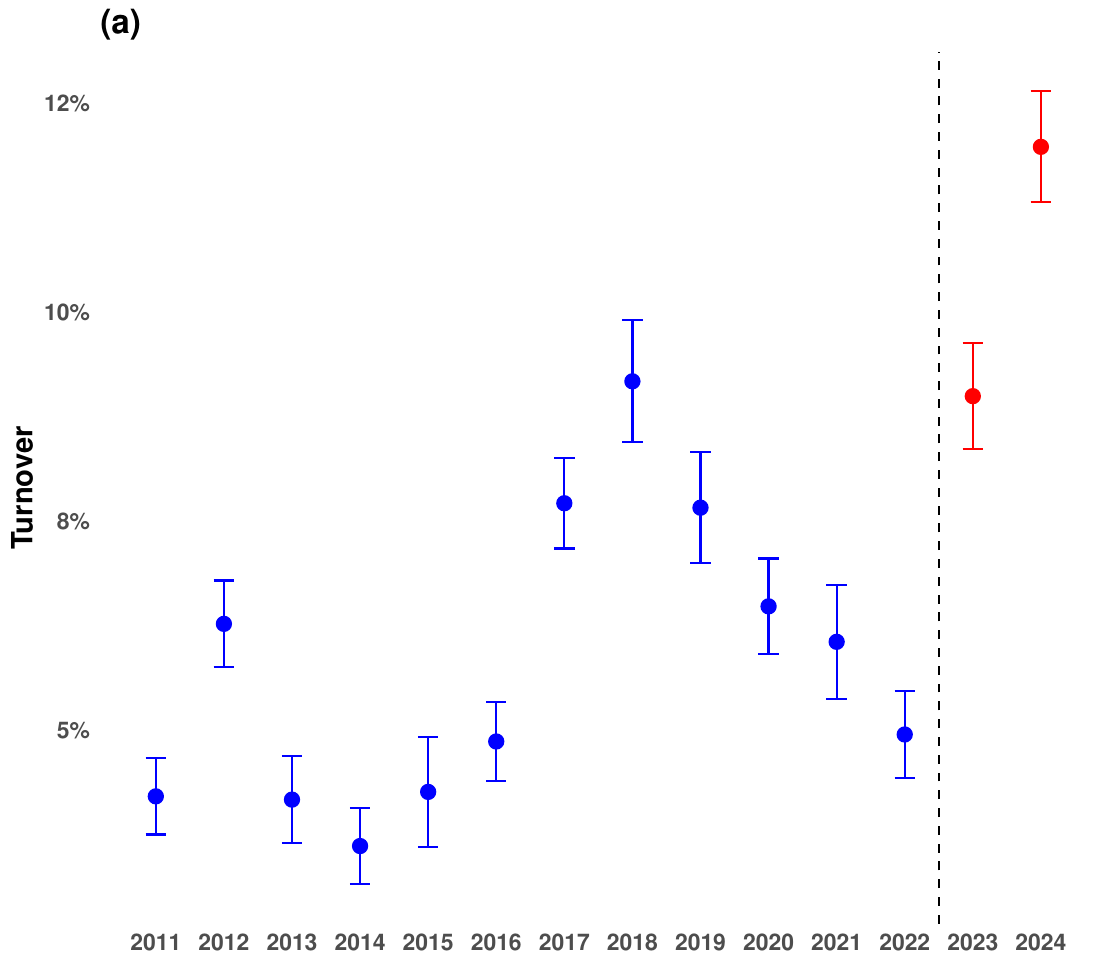}
\includegraphics[width=0.85\linewidth]{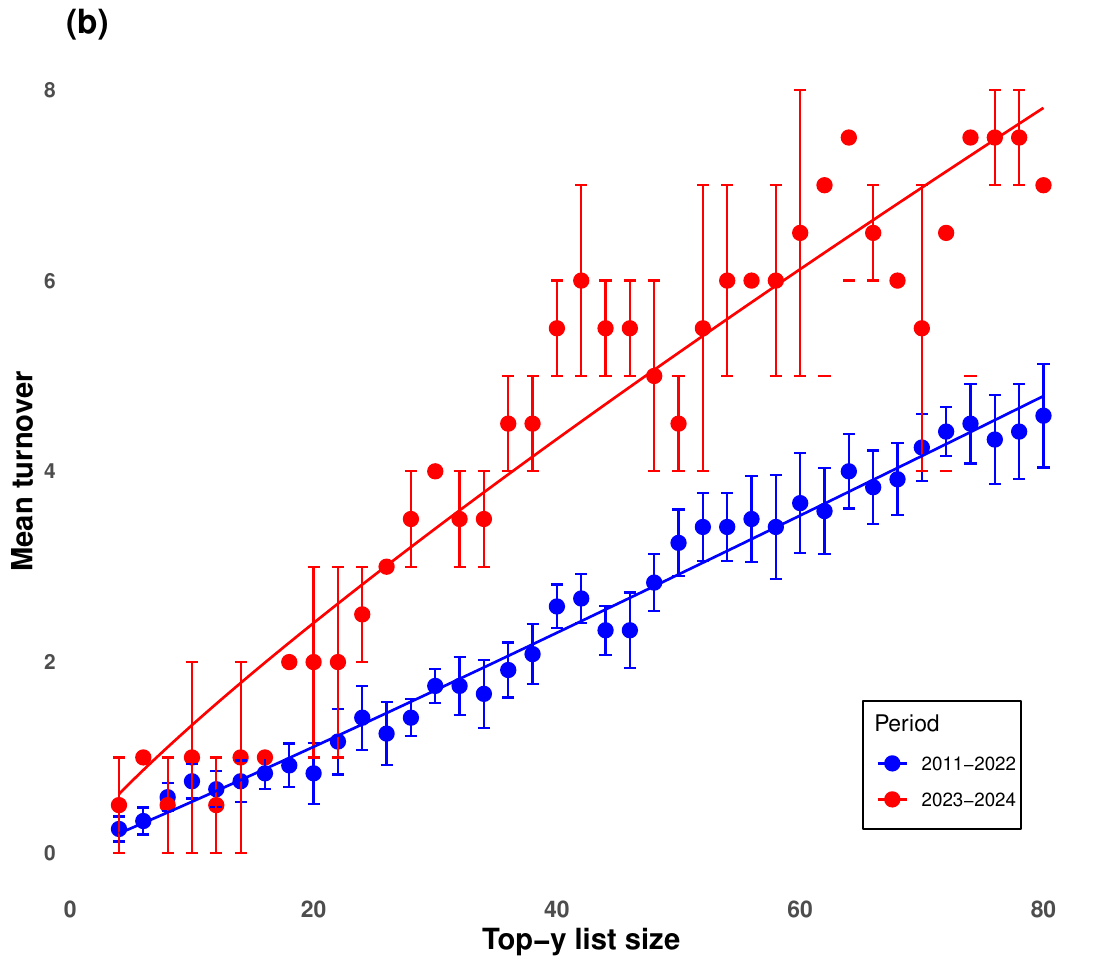}
\caption{Turnover (number of new words entering the top-$y$ list) among top-ranked content words in arXiv abstracts. (a) Mean annual fractional turnover in top-word lists, averaged across list sizes $y=4$ to $80$. The dashed line marks the transition from 2011–2022 (blue) to 2023–2024 (red). (b) Mean turnover versus list size $y$ for the same periods. Points show means $\pm$ s.e., and curves show fitted power-law relationships $z_y=ay^b$.}
    \label{fig:Turnover}
\end{figure}

Another revealing metric is top-$y$ turnover. Among top-ranked content words in arXiv abstracts, Figure \ref{fig:Turnover}a shows mean annual fractional turnover in top-word lists, averaged across list sizes $y=4$ to 80, for year-to-year transitions from 2011–2024. The dashed line marks 2023, after which turnover rises sharply. Figure \ref{fig:Turnover}b shows mean turnover (the number of new words entering the top-$y$ list) as a function of list size $y$, comparing transitions from 2011–2022 versus 2023–2024. Points show means $\pm$ s.e., and curves show fitted relationships $T(y)=ay^b$. Before 2023, turnover was less overall but increased slightly concave-up with list size ($b=1.15$), whereas in 2023–2024 the turnover is higher but slightly concave down ($b=0.969$). While both values of $b$ suggest modest conformity bias \cite{Acerbi&Bentley_2014}, turnover increased after 2023.

\clearpage
\section*{Discussion and Conclusion}

Here we analyze monthly changes in emergent lexical statistics of arXiv abstracts since 2014, with 2023 serving as a landmark year in which LLM use became widespread. We infer that the changes beginning in early 2023 are meaningful based both on the rapid adoption of generative AI tools after late 2022 and the abrupt rise in putative stylistic markers of LLM use.

Since early 2023, LLMs have contributed substantially to text in science \cite{Luo_etal_2024}, education and public discourse \cite{Brinkmann_etal_2023, Farrell_etal_2025}. Because these systems are trained on human language, their effects on the human–AI linguistic ecosystem are likely to be subtle \cite{Reinhart_etal_2025}. Consistent with this, arXiv abstracts show only modest changes in the exponents of Zipf’s law and Heaps’ law after 2023. Of greater interest are the more nuanced transitions that emerge at this point, including accelerated vocabulary growth, an increase in Heaps’ exponent, and higher turnover. Notably, the relationship between the composite LLM-associated style index and all three lexical measures—vocabulary size, Heaps’ exponent, and Zipf exponent—becomes flatter after 2023. Although these variables were already increasing over time, interaction models indicate that vocabulary growth after 2023 exceeds what would be expected from prior trends alone. Thus, current LLM contributions do not appear to reduce lexical diversity and may instead be associated with increased vocabulary richness, even as elevated turnover and changing scaling relationships point to more complex structural shifts.  These shifts in scaling relationships, together with the increased turnover, indicate changes in the dynamics of a mixed human–AI linguistic ecosystem.

If sustained, these patterns may have implications for longer-term dynamics, as humans increasingly offload cognitive processes—from human experts to search engines \cite{Sparrow_etal_2011}, and now to LLMs \cite{Fan_etal_2024}. More broadly, they point to a potential tension between short-term efficiency and long-term informational diversity. Emerging evidence that large reasoning models can exhibit performance collapse beyond complexity thresholds \cite{Shojaee_etal_2025} underscores the generality of this challenge. If diversity in human-generated and hybrid human–AI language declines over time, the consequences may extend beyond cultural evolution to the robustness of future AI systems trained on such data. Maintaining informational diversity may therefore be important for the resilience of both human knowledge systems and the AI models that increasingly depend on them.

\FloatBarrier
\section*{Material and Methods}
\small
\subsection*{Datasets}
We analyzed a large corpus of arXiv abstracts from January 2010 through September 2025 using a publicly available metadata dataset containing more than 2.7 million arXiv submissions, from which we analyzed first-version (v1) abstracts. These data are openly available at \url{https://www.kaggle.com/datasets/Cornell-University/arxiv}. Each abstract is identified by an arXiv identifier encoding submission year and month. Abstracts were tokenized into one token per row without stopword removal, stemming, or lemmatization in order to preserve full lexical distributions. Word boundaries were defined using the regular expression \texttt{\textbackslash b\textbackslash w+\textbackslash b}. All scaling analyses were conducted on these raw lexical tokens.

To benchmark LLM outputs against human-authored text, we applied the same Zipf and Heaps analyses to the English Human–ChatGPT Comparison Corpus (HC3) \cite{Gao_etal_2023, Su&Wu_2024}, which comprises ChatGPT responses and three human-authored subsets. 

\subsection*{Analyses}
We constructed a composite LLM-associated style index based on monthly frequencies of dash markers (en-dashes, em-dashes, double hyphens and triple hyphens) and selected lexical markers--“significant,” “crucial,” and “showcase”\cite{Geng&Trotta_2024}-- in arXiv abstracts from 2014 onward. The index is a descriptive proxy for stylistic change, if not a direct measure of LLM-generated content. For each month, we computed the normalized rate per 1000 words for each feature. These monthly rates are then standardized across the full time series as $z = \frac{x - \mu}{\sigma}$
where $\mu$ and $\sigma$ denote the mean and standard deviation of the monthly series for each feature. The LLM-associated style index is defined as the sum of these standardized components:
\begin{equation}
\text{LLM\_index} = z_{\text{dash}} + z_{\text{significant}} + z_{\text{crucial}} + z_{\text{showcase}}.
\end{equation}
This index captures the joint prevalence of stylistic features that increase markedly after early 2023 and is used as a descriptive measure of temporal changes in lexical patterns. 

Zipf’s law characterizes the rank–frequency distribution of words. For each year and month, tokens were pooled, word frequencies were computed, and ranks assigned in descending order. The exponent $\alpha$ was estimated by maximum likelihood under a finite discrete Zipf distribution, which is more appropriate for ranked count data than ordinary least-squares fits on log–log scales. Specifically, we assume
$P(r) \propto r^{-\alpha}$, where $r$ denotes rank. For a finite vocabulary of size $N$, the likelihood is
\begin{equation}
L(\alpha) = \prod_i \frac{r_i^{-\alpha}}{H_{N,\alpha}},
\end{equation}
where
\begin{equation}
H_{N,\alpha} = \sum_{r=1}^{N} r^{-\alpha}
\end{equation}
is the generalized harmonic number. The log-likelihood is then
\begin{equation}
\ell(\alpha)
=
-\alpha \sum_i \log r_i
-
n \log H_{N,\alpha},
\end{equation}
which was maximized numerically to obtain $\hat{\alpha}$.

Heaps’ law was modeled as $V(N)=aN^b$, where $N$ is cumulative token count and $V(N)$ is the number of unique word types. For arXiv abstracts, vocabulary growth was computed sequentially in increments of 20,000 tokens, yielding sufficient points for reliable estimation while smoothing short-range fluctuations. Within each increment, newly observed types were tracked using a hash table, and parameters were estimated by nonlinear least squares. Only corpora with at least three increments were retained. For HC3, Heaps’ exponents were estimated from repeated random subsets of 100–1000 responses, repeated 1,000 times per subset to obtain confidence intervals.

To quantify lexical turnover through time, we examined changes in yearly top-ranked vocabularies in arXiv abstracts after removing standard English stopwords and a small set of high-frequency scientific boilerplate terms (e.g. “method,” “model,” “data”). For each year from 2010–2024, words were ranked by frequency and top-$y$ lists were constructed for list sizes $y = 4,6,\dots,80$. For each adjacent year pair, turnover was defined as the number of words appearing in the top-y list at year $t+1$ that were absent from the top-$y$ list at year $t$, with a fractional turnover measure obtained by dividing by $y$. Mean turnover as a function of list size was compared between pre-LLM transitions (2011–2022) and post-2023 transitions (2023–2024), and fitted with a power-law relation $T(y)=ay^b$ using nonlinear least squares. To assess temporal changes in turnover independent of list size, we also averaged fractional turnover across all $y$ values for each annual transition and examined its trajectory through time.

\paragraph{Acknowledgments:} {\footnotesize S.V. was supported by the Spanish Ministry of Science and Innovation through the State Research Agency (AEI), grant PID2024-162055NB-I00/MICIU/AEI/10.13039/501100011033/FEDER. B.V. and S.V. were supported by grant PCI2022-132936 funded by MCIN/AEI/10.13039/501100011033 and by the European Union NextGenerationEU/PRTR.}

\vspace{1cm}

\FloatBarrier
\begingroup
\footnotesize
\setstretch{0.88}

\endgroup

\end{document}